\documentclass[structabstract]{aa}
\usepackage{graphicx}
\usepackage{txfonts}
\usepackage{natbib}
\bibpunct{(}{)}{;}{a}{}{,}

\begin{document}

\title{S2DFS: Analysis of temporal changes of drifting subpulses}
\subtitle{}

\author{M.~Serylak\inst{1,2}, B.~W.~Stappers\inst{3} \and P.~Weltevrede\inst{4} }

\institute{Netherlands Institute for Radio Astronomy, P.O. Box 2, 7990 AA Dwingeloo, The Netherlands\\
           \email{serylak@astron.nl}
           \and
           Astronomical Institute 'Anton Pannekoek', University of Amsterdam, P.O. Box 94249, 1090 GE Amsterdam, The Netherlands
           \and
           Jodrell Bank Centre for Astrophysics, Alan Turing Building, University of Manchester, Manchester M13 9PL\\
           \email{ben.stappers@manchester.ac.uk}
           \and
           Australia Telescope National Facility, CSIRO, P.O. Box 76, Epping, NSW 1710, Australia\\
           \email{patrick.weltevrede@atnf.csiro.au}
          }


\date{Received ; accepted}

\abstract
{We introduce a new technique, called the Sliding Two-Dimensional Fluctuation Spectrum, used for detecting and characterising the temporal changes of drifting subpulses from radio pulsars. The method was tested using simulated data as well as archived observations made with the Westerbork Synthesis Radio Telescope (WSRT) at wavelengths of 92 and 21 cm.}
{The drifting subpulse phenomenon is a well known property of radio pulsars. It has been studied extensively since their discovery, however the properties of the temporal behaviour of drifting subpulses are not fully explored. The drifting can also be non-coherent and the presence of effects like nulling or drift rate changing can mask the drifting behaviour of the pulsar. The S2DFS is a robust method for investigating this phenomenon and by introducing it we aim to expand our knowledge of the temporal drifting subpulse properties.}
{Our new analysis method uses horizonally collapsed fluctuation spectra obtained with the Two-Dimensional Fluctuation Spectrum (2DFS) method. Stacking the collapsed spectra obtained in a 256 pulse window which slides by a pulse at a time produces a map of the collapsed fluctuation spectrum (S2DFS plot). By analysing the maps one can easily determine the presence of any temporal drift changes.}
{Simulated data showed that the technique can reveal the presence of any temporal changes in drift behaviour like mode changing or nulling. We have also analysed data of three pulsars, PSRs B0031$-$07, B1819$-$22 and B1944$+$17, which were selected based on the quality of the data and their known drift properties. All three sources are known to exhibit mode changes which could easily be seen in the S2DFS.}
{The results from the analysis of the data sets used in this paper have shown that the S2DFS method is robust and complimentary to the 2DFS method in detecting and characterising the temporal changes in drifting subpulses from radio pulsars.}


\maketitle

\section{Introduction}

\begin{figure}
\resizebox{\hsize}{!}{\includegraphics[angle=270]{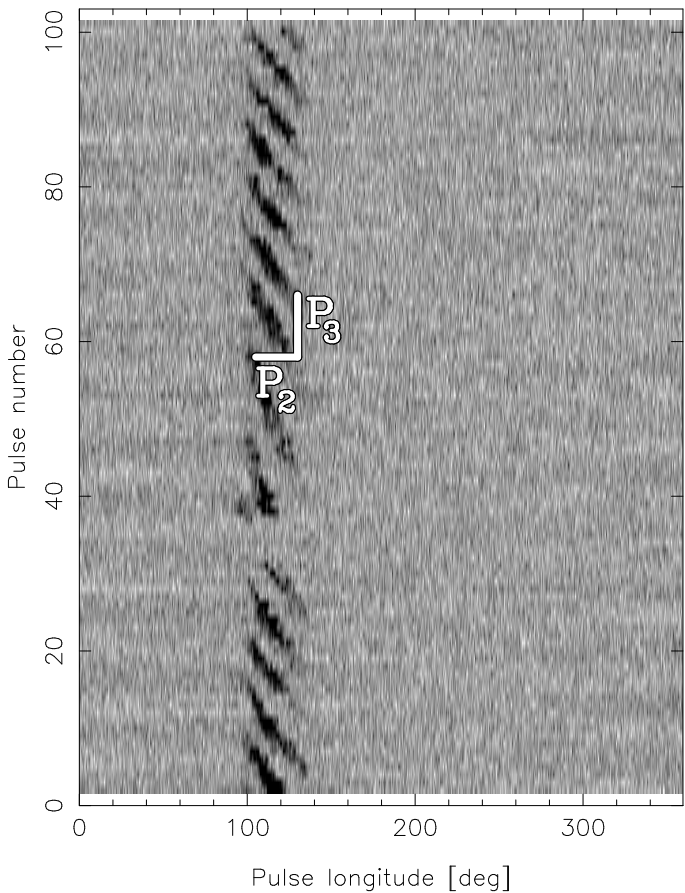}}
\caption{An example of sequence of one hundred successive single pulses of PSR B0031$-$07 observed at 92 cm. The subpulses appear earlier with increasing pulse number and are arranged into so-called drift bands. Two successive drift bands are vertically separated by $P_{3}$ and horizontally by $P_{2}$.}
\label{fig:pulse_stack}
\end{figure}

In the years following the serendipitous discovery of pulsars \citep{hbp+68}, the results from analysis of long sequences of pulses revealed characteristic ``second periodic pulsations'' seen in the plots of the subpulse intensity \citep{dc68}. This phenomenon, called ``class 2 period'' (denoted by $P_{2}$, see Fig.~\ref{fig:pulse_stack}), expressed itself as a constant change of the phase of successively appearing subpulses within a fixed longitude range. In order to investigate this new phenomenon, numerous studies were conducted \citep[e.g.][]{lc68, tjh69, lan69, sspw70, col70a}. The authors performed fluctuation-spectral analysis using Fourier techniques which revealed another type of characteristic frequency at which a significant amount of power was present in the fluctuation spectra of pulsars and provided evidence of a periodic modulation of pulse intensities. These intensity modulations were described by a pattern period $P_{3}$ (see Fig.~\ref{fig:pulse_stack}), which is the time between the successive appearance of subpulses at a particular pulse phase, expressed in pulsar periods, $P_{0}$ \citep{sspw70}. Although very useful, these techniques had their limitations, and a better technique was needed.

The work of \citet{bac70b, bac70a, bac73} and \citet{brc75} systematised the knowledge and introduced a better fluctuation-spectral analysis technique by applying the Discrete Fourier Transform (DFT) to regions narrower than the pulse width, that are ``Pulse Phase Boxes'' and which insight into the properties of subpulse modulation within a pulse profile. This technique, known as the Longitude-Resolved Fluctuation Spectrum (LRFS), became very widely used in the analysis of the subpulse drift properties of pulsars. Apart from discovering and characterising the $P_{3}$ values, the LRFS method could not resolve information about subpulses drift (i.e. $P_{2}$ value). This insformation was obtained by applying the Fourier transform to a series of pulses, but it was often replaced by a visual inspection of high signal-to-noise ratio (S/N) observations. \citet{es02} presented a new technique, the Two-Dimensional Fluctuation Spectrum (2DFS) which could easily determine the $P_{2}$ value. The 2DFS is an extension of the LRFS, in a sense, where, as well as applying the DFT along vertical lines of constant phase in the pulse stack (the result of which gives the LRFS), a second DFT is applied to each row of the complex version of the LRFS, thus producing the two-dimensional power spectrum. While this results in the loss of pulse longitude information, it provides the $P_{2}$ value. The LRFS and 2DFS techniques have proven to be robust and effective for discovering and characterising subpulse drifting from radio pulsars as shown in the recent work by \citet{wes06, wse07}. The Harmonic-Resolved Fluctuation Spectrum technique, called HRFS, developed by \citet{dr01}, can also be used for subpulse drifting analysis and is mathematically equivalent to the 2DFS.

In their work, Weltevrede and collaborators examined a large sample of pulsars observed at two frequencies in order to study their subpulse modulation properties. A large fraction of all pulsars (about 40\% of a total sample of 185 sources) were reported to exhibit the drifting subpulse phenomenon. Those pulsars whose $P_{3}$ values were well defined have been classified as coherent drifters, while the sources for which $P_{3}$ could not be well defined, were assigned to the diffuse class. Due to various reasons like low S/N, interstellar scintillation effects, severe radio frequency interference during observations or conditions intrinsic to the observed source, not all pulsars in the sample were observed to exhibit the drifting subpulse phenomenon \citep{wes06}.

Let us now consider the pulsar intrinsic effects: The first intrinsic property that might prevent the detection of drifting is the viewing geometry. In this case the line of sight cuts the magnetic pole centrally and only longitude stationary subpulse modulation is expected. The second intrinsic property can be assigned to the refractive properties of the pulsar magnetosphere leading to a distorted pulsar signal. We will now focus on the other effects potentially responsible for the lack of drifting: nulling and mode changing.

Nulling characterises itself as a sudden diminishing or cessation of pulse energy below the level of detection for a certain amount of time, and was first reported by \citet{bac70} in the observations of PSRs B0826$+$06, B1133$+$16, B1237$+$25 and B1929$+$10. In Fig.~\ref{fig:pulse_stack} one can see an example of a null between pulses 31 and 38 in the pulse stack of PSR B0031$-$07. Studies which followed this discovery \citep[e.g.][]{rit76, ran86, big92, wmj07} have shown that nulling can occur on random timescales while the duration can vary between one or more periods to longer timescales of hours to days. The ``Nulling Fraction'', which is the percentage of time for which the pulsar nulls ranges from almost zero (for the Vela pulsar) up to 93 \% (for PSR J1502$-$5653) as shown by \citet{wmj07}. It is also worth noting that although nulling is a broadband effect, nulls do not always occur simultaneously at different simultaneously observed frequencies \citep{dls+84, bgk+07}.

Pulsars are known to form extremely stable profiles made after averaging just a few hundred pulses \citep{hmt75, rr95}, but in some cases they change the shape of their average profiles as first observed in PSR B1237$+$25 and reported by \citet{bac70a}. This behaviour, called mode changing, is known to happen in about 12 pulsars, mostly with multicomponent average profiles \citep{bmsh82, ran86, wmj07}. The mode changing happens suddenly, it can include two or more modes and can last from a few to thousands of pulse periods. It is also important to note that mode changing is closely connected with the nulling and drifting subpulse phenomena. For example, \citet{htt70} have first shown that PSR B0031$-$07 changes its drift rate in each of its modes, while PSR B0809$+$74 after many or all nulls, emits in a mode different from the normal one \citep{LKR+02}.

The presence of the above mentioned effects can disrupt the subpulse drift of the pulsars and change their subpulse modulation periodicities which, depending on the magnitude of the effects and the S/N of the observation, can influence the fluctuation analysis. We introduce an extension of the 2DFS technique called the S2DFS for discovering and characterising the temporal changes of drifting subpulses from radio pulsars. We begin with a more detailed description of the 2DFS technique for analysing subpulse modulation phenomena. We then introduce the new technique and apply it to the simulated and archival data of a few well known pulsars observed with the Westerbork Synthesis Radio Telescope (WSRT). Finally, we discuss the implications of the application of our method to pulsars with unstable drifting behaviour, multiple drift modes or those affected by the nulling phenomenon.

\section{Methods of analysis}

\subsection{2DFS}

In order to analyse the subpulse modulation phenomena using the methods described in this work, for each data set, a so-called pulse stack has to be formed. To create a pulse stack, a de-dispersed time series is transformed into a longitude and time-resolved representation of the pulsar signal. This is done by stacking time samples from consecutive pulses, according to their phases, as calculated using an ephemeris and the TEMPO software package\footnote{http://www.atnf.csiro.au/research/pulsar/tempo}. Fig.~\ref{fig:pulse_stack} presents an example of a pulse stack of one hundred pulses from PSR B0031$-$07 plotted one above the other, where the pulse longitude is plotted on the horizontal axis and the pulse number on the vertical axis. One can easily see that the subpulses appear earlier with increasing pulse number and are arranged into so-called drift bands.

\begin{figure*}
\centering
\includegraphics[width=137mm]{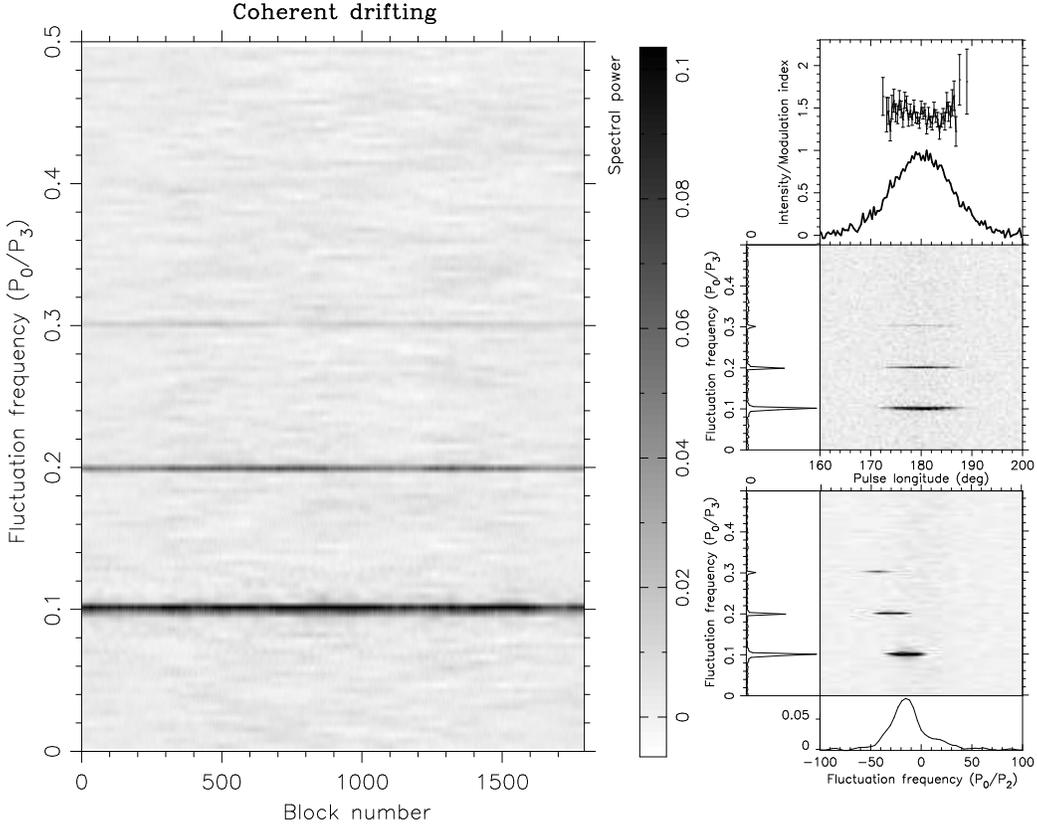}
\caption{The analysis results shown for a simulated pulse stack assuming the coherent drifting scenario. {\bf Left:} The S2DFS. The vertical axis is given in $P_{0}/P_{3}$ (cpp) and corresponds to the vertical separation of drift bands. The horizontal axis is given in blocks, where each block is a curve obtained from the vertically collapsed 2DFS where the DFT window of 256 pulses ``slides'' along the pulse stack. Subsequently, all curves are stacked together producing maps of the collapsed fluctuation spectra. {\bf Right:} The fluctuation analysis results. The top panel shows the integrated pulse profile (solid line), the longitude-resolved modulation index, LRMI (solid line with error bars). The middle panel shows the LRFS where the ordinate of the resulting spectrum is given in cycles per period (cpp) which corresponds to $P_{0}/P_{3}$ ($P_{0}$ is the pulsar period and $P_{3}$ denotes the vertical separation of drift bands) and the abscissa denote pulse phase given in degrees. The bottom panel represents the 2DFS where the ordinate is the same as the LRFS but the abscissa is in cpp ($P_{0}/P_{2}$) ($P_{2}$ denotes the horizontal separation of drift bands). The greyscale intensity of the 2DFS corresponds to the spectral power. The presence of a significant spectral feature with a negative or positive value of $P_{2}$ indicates that the subpulses appear with a preferred direction of drifting subpulses with periodicity of $P_{0}$ pulsar periods. The side panel corresponds to the horizontally (left-hand panel) and vertically (bottom panel) integrated spectrum between the dashed lines.}
\label{fig:coherent_drifting}
\end{figure*}

The basic method to estimate the presence of subpulse modulation is to calculate the longitude-resolved modulation index (LRMI; \citealt{es02}). The LRMI is the measure of the factor by which the intensity varies from pulse to pulse for each pulse longitude. In the right-hand plot in Fig.~\ref{fig:coherent_drifting} the average profile and the LRMI are shown in the top panel for a simulated pulse stack. The solid line corresponds to the average profile, which is normalised to the peak intensity, while the solid line with error bars denotes the LRMI values. We only show LRMI values that are detected with a significance more than $3\sigma$.

Although the LRMI is the first evidence of subpulse modulation, it does not indicate whether this modulation is systematic or lacks organisation. To obtain such information the LRFS has to be calculated. This is done by dividing the pulse stack into blocks of $2^{n}$ pulses and applying a DFT at each pulse longitude bin in each block (for details of the analysis we refer to \citealt{es02, es03}). The final spectrum is produced by averaging each of the longitude-resolved spectrum in each block. If the pulsar exhibits subpulses which are periodically modulated, then a region, so-called feature, of enhanced power will be visible as a dark region in the greyscale of the power in the LRFS. The vertical position of the feature is given in cycles per period (cpp) and corresponds to a period of modulation, $P_{3}$ expressed in pulsar periods $P_{0}$, while the horizontal position denotes the pulse longitude position at which the modulation occurs. In the right-hand plots in Fig.~\ref{fig:coherent_drifting} the LRFS of the simulated pulse stack is shown below the top panel.

While the LRMI allows one to detect the presence of subpulse modulation, and the LRFS determines whether subpulse modulation is disorganised or periodic, the information about whether the subpulses are drifting and with what rate is not resolved. Such information can be acquired by applying the 2DFS. The procedure is similar to the calculation of the LRFS where, except now after applying the DFT to the pulse stack along the vertical lines of constant phase, the DFT is applied again, but now across each row of the complex LRFS spectrum. In the right-hand plot in Fig.~\ref{fig:coherent_drifting} the 2DFS for simulated pulse stack is plotted in the panel below the LRFS plot. The vertical axis of the resulting spectrum is the same as in the LRFS, but the horizontal axis is now given in cycles per period and corresponds to a horizontal separation of the drift bands ($P_{2}$ expressed in pulse periods). If the feature is significantly offset from the vertical axis ($P_{0}/P_{2} \neq 0$) it means that subpulses drift in a certain longitude range have a preferred drift direction. A negative or positive value of $P_{2}$ denotes the direction of the drift, i.e. whether subpulses drift towards the leading or trailing edge of the pulsar's average profile respectively. The power of the 2DFS is vertically (between dashed lines) and horizontally integrated, resulting in the plots in the side and bottom panels, which are used for better understanding the structure of the feature. To measure values of $P_{2}$ and $P_{3}$ one has to calculate the centroid of a rectangular region in the 2DFS containing the feature (Eq.~6; \citealt{wes06}), while errors for those values are estimated by applying the same procedure to an equivalent rectangular region containing only noise.

\begin{figure*}
\centering
\includegraphics[width=137mm]{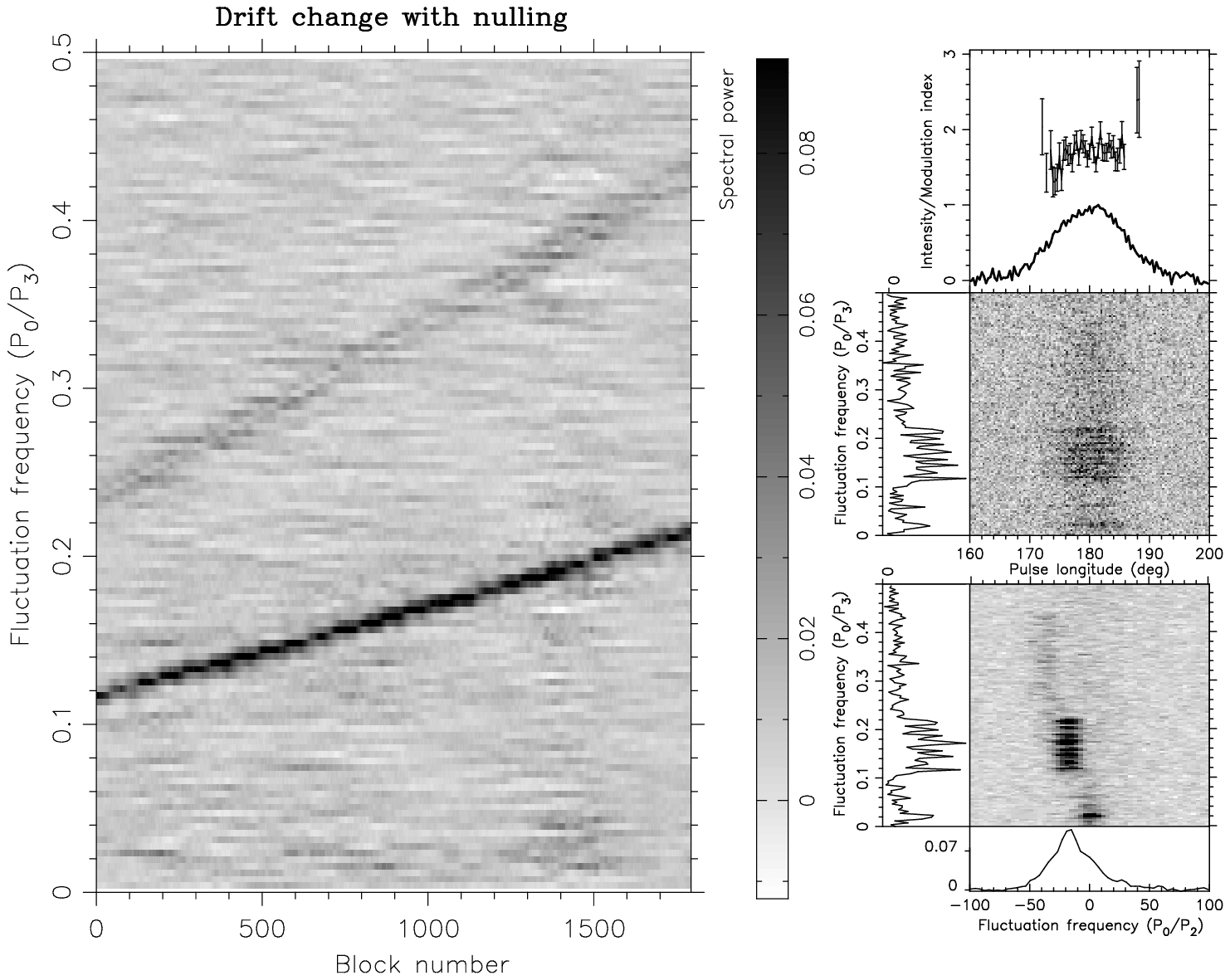}
\caption{The analysis results shown for a simulated pulse stack assuming a drift rate change in combination with nulling. For the explanation of the plots we refer to Fig.~\ref{fig:coherent_drifting} and the text.}
\label{fig:drift_change_with_nulling}
\end{figure*}

\subsection{S2DFS}

The 2DFS method is a very effective tool for subpulse drift detection and analysis. Its advantage over other methods lies in a process of averaging multiple fluctuation spectra, obtained from consecutive blocks of pulses, into a final fluctuation spectrum. This can overcome the problem of low S/N observations. However, for pulsars which are affected by phenomena like drift mode changing, nulling or other effects, the results of the aforementioned spectral analysis do not show pure drift features. The process of averaging also results in a loss of temporal information, making the detection and characterisation of any time-related changes very difficult. In this section we propose an extension to the 2DFS technique which can resolve temporal changes of drift rates of subpulses.

We call this method the Sliding Two-Dimensional Fluctuation Spectrum (S2DFS). The spectrum is obtained in the same way as in the 2DFS but instead of dividing a pulse stack, which consists of N pulses, into consecutive blocks of equal length, the 2DFS is applied to a block of M pulses, and a spectrum is calculated. The spectrum is collapsed horizontally producing a one-dimensional curve (see for example the left-hand side panel of the 2DFS plot in Fig.~\ref{fig:coherent_drifting}, right-hand plot). Subsequently, the DFT ``window'' is shifted by one pulse and the whole process is repeated. The application of the sliding DFT window to the pulse stack results in $N-M+1$ curves, which are stacked together to produce a so-called ``map'' of the collapsed fluctuation spectra (see Fig.~\ref{fig:coherent_drifting}, left-hand plot). Changing the position of the DFT window and calculating the 2DFS allows one to resolve any changes of drift over relatively short timescales. The choice of the length of the DFT window is crucial for the resolution and sensitivity of the maps. If the size of sliding window, and hence the length of the Fourier transform, is too small, the resulting details will not be resolved due to the insufficient resolution in $P_{0}/P_{3}$ space, but if the size of the window is too large then the method will not be sensitive enough, and any short-lasting events visible in the pulse stack will not be sufficiently prominent in the maps due to the reduced S/N per spectral bin. In our work we have decided to use a DFT window size of 256 pulses and as will be shown later it has proven to be a good compromise between resolution and sensitivity. We note that in order to generate the aforementioned maps of the collapsed fluctuation spectra one can also use the LRFS method. We have decided to use 2DFS-based procedure, because collapsing the fluctuation spectra vertically can be used to obtain the information regarding changes of $P_{2}$ values. However, the maps of vertically collapsed fluctuation spectra, holding information about the $P_{2}$ temporal changes, usually leads to results comparable to those from the 2DFS method alone.

\section{Simulated data}

To test the new technique of subpulse modulation analysis we have generated a number of artificial pulse sequences. We used an emission model where the emission is generated by a rotating ``carousel'' of discharges, or so-called ``sparks'', which circulate around the magnetic axis due to an ${\bf E} \times {\bf B}$ drift \citep{rs75}. We know there are many problems with the carousel model, but it proves to be sufficient for generating pulse stacks for testing our new algorithm, and we are not at this stage, concerned about the origin of the mechanism that produces the drifting subpulses. We have chosen the set of elementary model parameters used to generate the pulse stacks to be similar to the pulsar PSR B0809$+$74. These parameters are kept identical in all the pulse stacks in order to make comparison between the data sets easier. Specifically, we used the following pulsar geometry: the angle between rotation and magnetic axes of the pulsar, $\alpha = 9$ deg and the angle between rotation axis and the line of sight, $\zeta = 13.5$ deg. The pulse stacks consist of a total of 2048 pulses with a period of $P_{0} = 1$ s. The period value is not used in our calculations and has no importance for the analysis presented in this paper. We set the number of sparks $N = 5$ and we have chosen a drift rate value $P_{3} = 10\ P_{0}$. Each subpulse has a full width at half-maximum (FWHM) of $w_{50} = 5$ deg. The pulse period is divided into one thousand longitude bins and the average pulse profile has a FWHM of $w_{50} = 12$ deg and $S/N = 100$.

The pulse sequences have been arranged into a few types of emission scenarios to investigate the effects of nulling, changing drift rates and mode changing on the S2DFS. In addition, to each emission scenario we introduced several effects which alter the fluctuation analysis results. The first effect was pulse-to-pulse intensity modulation with a normally distributed range of subpulse amplitudes. This makes the modulation pattern less pure, hence it broadens the feature in the spectra, making it more difficult to detect. The second effect was a steady emission component with a FWHM of $w_{50} = 12$ deg and an amplitude of 10 \% of the average pulse profile. In some cases we also combined both the steady emission component and pulse-to-pulse modulation thogether.

\subsection{Coherent drifting}

\begin{table*}
\caption{The details of the analysed pulsars. The last two columns hold results presented in the work of \citet{wes06, wse07}.}
\label{table:pulsar_details}
\centering
\begin{tabular}{c c c c c c c c c}
\noalign{\smallskip}
\hline
\hline
\noalign{\smallskip}
Name & Date           & $\nu$ & $P_{0}$ & $N_{pulses}$ & $P_{3}$   & $P_{2}$ \\
PSR  & [MJD/dd.mm.yy] & [MHZ] & [s]     &              & [$P_{0}$] & [deg]   \\
\noalign{\smallskip}
\hline
\noalign{\smallskip}
B0031$-$07 & 51780/24.08.2000 &  328 & 0.9430 & 7666 &  $6.7\pm 0.1$ & $-21.9^{+0.2}_{-0.1}$ \\
\noalign{\smallskip}
           & 53239/22.08.2004 & 1380 &        & 1204 &  $8.3\pm 0.3$ &      $-40^{+2}_{-50}$ \\
\noalign{\smallskip}
B1819$-$22 & 53182/26.06.2004 &  328 & 1.8743 & 1095 & $16.9\pm 0.6$ &       $-17^{+2}_{-6}$ \\
\noalign{\smallskip}
           & 53188/02.07.2004 & 1380 &        & 1096 &  $9.8\pm 0.6$ &  $-9.1^{+0.1}_{-1.0}$ \\
\noalign{\smallskip}
B1944+17   & 52706/08.03.2003 &  328 & 0.4406 & 3990 & $13.5\pm 0.9$ &      $-30^{+5}_{-10}$ \\
\noalign{\smallskip}
           & 51977/09.03.2001 & 1380 &        & 2056 &   $13\pm 0.5$ & $-11.7^{+0.3}_{-0.3}$ \\
\noalign{\smallskip}
\hline
\end{tabular}
\end{table*}

The first emission scenario is a simulation of coherently drifting subpulses with a chosen drift value of $P_{3} = 10\ P_{0}$. The 2DFS of this scenario (see Fig.~\ref{fig:coherent_drifting}, right-hand plot) showed a very strong and narrow ($< 0.05$ cpp) feature, at 0.1 cpp, corresponding to a $P_{3} = 10\ P_{0}$ with $P_{2} = 20$ cpp (corresponding to $360$ deg $/~20$ cpp $= 18$ deg pulse longitude). As expected, as there were no distorting effects, the S2DFS plot shows no change of drift rate over time in any of the simulated sequences.

\begin{figure*}
\centering
\includegraphics[width=137mm]{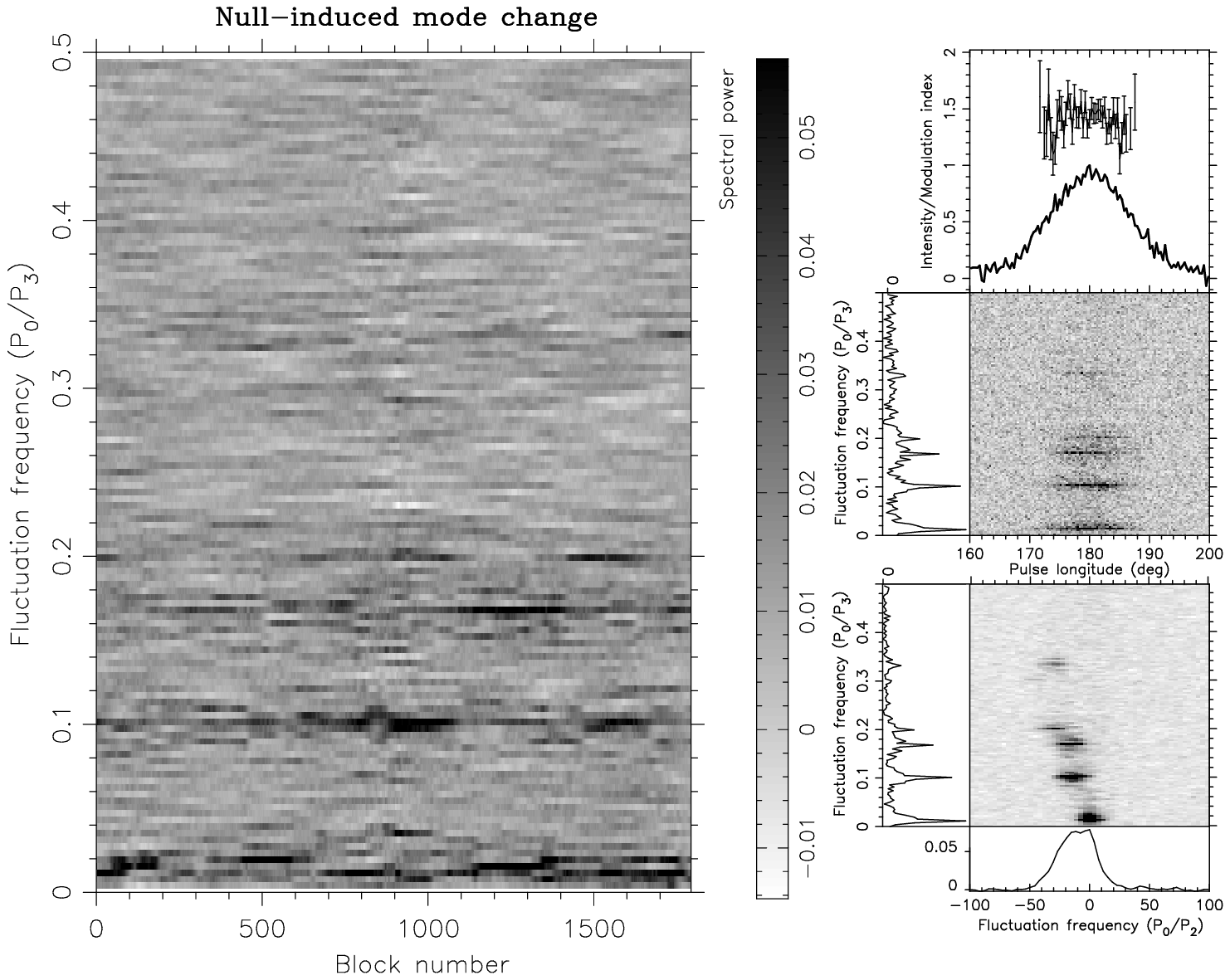}
\caption{The analysis results shown for a simulated pulse stack assuming null-induced mode changes. For the explanation of the plots we refer to Fig.~\ref{fig:coherent_drifting} and the text.}
\label{fig:null-induced_mode_change}
\end{figure*}

The introduction of pulse-to-pulse modulation, a steady emission component or both effects combined did not strongly affect the spectral results (not shown). However, after introducing pulse-to-pulse modulation we noted the presence of a very weak, low-frequency feature around $P_{0}/P_{3}$ and $P_{0}/P_{2} = 0$ in the 2DFS plots of the analysed pulse stacks. We also note that despite the relatively low S/N all of the above spectral results showed second and third harmonics of the main drift feature in this scenario (see Fig.~\ref{fig:coherent_drifting}). We interpret this as being due to the coherence and non-sinusoidal shape of the drift bands as discussed in \citet{es02}.

In the next scenario, we introduced nulling to the pulse stack. The null lengths and positions were randomly generated with a minimum nulling length of 5 pulses and total nulling fraction, NF = $22.9 \%$. As in the coherent drifting scenario, the 2DFS showed a strong and narrow feature with $P_{3} = 10\ P_{0}$. We also note the presence of a low-frequency feature in the 2DFS plots (not shown), which we identify as being due to the presence of nulls. This feature is similar to the feature corresponding to the presence of pulse-to-pulse intensity modulation. We note that nulling, due to its timescales and occurrence rates, can be seen as an extreme form of intensity modulation, hence it is seen as a feature close to the $P_{0}/P_{3} = 0$ border. Also, due to the nulling only the second harmonic was visible in both the 2DFS and S2DFS plots in this scenario due to the weaker coherence of the modulation pattern.

\subsection{Drift rate change}

The next two scenarios included a constant change of drift rate, where the initial value of $P_{3} = 8.5\ P_{0}$ changes to a value of $P_{3} = 4.5\ P_{0}$ after 2048 periods, and in the case of the second scenario nulling was also introduced with a nulling fraction of NF = $22.9 \%$. In both cases the 2DFS plots showed eight closely located drift features spanning 0.1 cpp, along the $P_{0}/P_{3}$ axis and in the second scenario a low-frequency feature as in the coherent drift with nulling scenario is seen (see Fig.~\ref{fig:drift_change_with_nulling}, right-hand plot). These unusual looking spectra are caused by the constant change of $P_{3}$ and a property by which the 2DFS is calculated. This property is related to the process of calculating and averaging multiple spectra to obtain the final spectrum. When the pulse stack is divided into blocks of pulses and the DFT is applied to each consecutive block, each block's spectrum shows spectral features offset in the vertical direction from the spectrum of the first block. Those features correspond to the different values of the changing $P_{3}$. The resulting block spectra were averaged together to form the final 2DFS plot. What determines the number of spectral features in the 2DFS is the number of pulses in the pulse stack and the size of the block of pulses used in the DFT.

In the case of the low S/N data it is very difficult to inspect the pulse stack by eye, making the spectral analysis the only reliable method to determine the initial drift rate or the speed of its change. However, from the 2DFS alone one is not able to determine such drift rate variations. The S2DFS technique allows one to circumvent this problem as the process of averaging is replaced by a moving DFT window. In the left-hand plot of Fig.~\ref{fig:drift_change_with_nulling} an example S2DFS plot is presented where one can easily see the ``tracks'' corresponding to the changing value of the main drift feature and its second harmonic. By analysing the S2DFS plot it is quite straightforward to determine the initial value of $P_{3}$ and its rate of change. Also inspection of the S2DFS plot can reveal the presence of nulling. This can be seen in the S2DFS plot (Fig.~\ref{fig:coherent_drifting}, left-hand plot) between blocks 100 to 200 or 1400 to 1500 at 0.02 cpp.

\subsection{Null-induced mode change}

\begin{figure*}
\resizebox{\hsize}{!}{\includegraphics{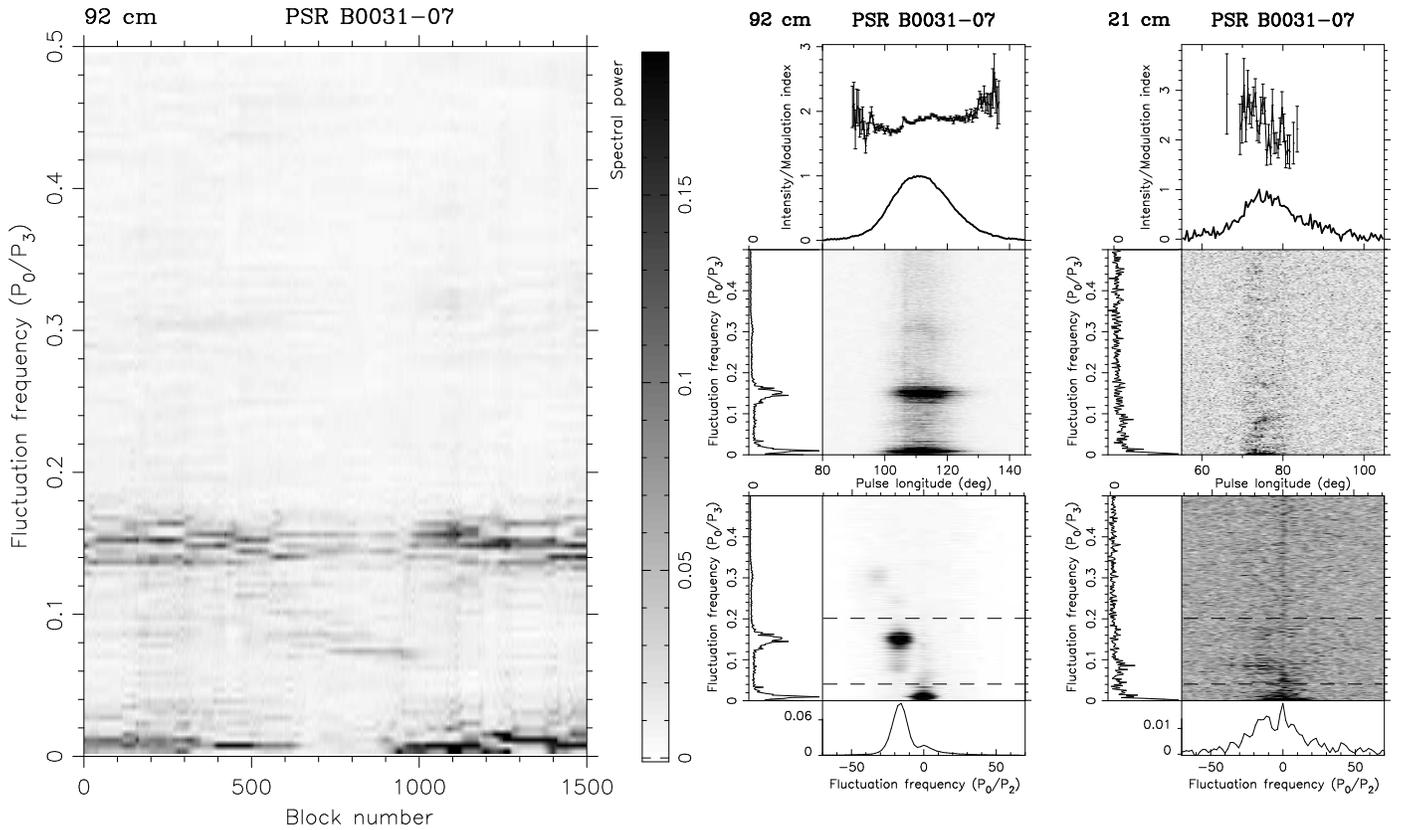}}
\caption{The analysis results shown for PSR B0031$-$07. For the explanation of the plots we refer to Fig.~\ref{fig:coherent_drifting} and the text.}
\label{fig:0031}
\end{figure*}

For the final simulated pulse sequence we investigated the null-induced mode change scenario. The generated pulse stack had two drift modes, mode ``A'' with $P_{3} = 10\ P_{0}$ s, and mode ``B'', $P_{3} = 6\ P_{0}$ s, where the numbers of pulses in each mode was randomly chosen. The mode change was induced after a sequence of null pulses. The null lengths were also randomly generated. The minimal nulling length which triggered the change of the drift modes was set to eight periods, the overall nulling fraction was, NF = $32.4 \%$ and the calculated numbers of pulses in the drift modes ``A'' and ``B'' contained 797 and 587 pulses respectively. The return to the preceeding drift mode was executed abruptly without recovery time.

Fig.~\ref{fig:null-induced_mode_change} (bottom right-hand plot) presents the 2DFS of this case where a pulse-to-pulse modulation and a steady emission component were also introduced to the pulse stack. In the spectrum one can easily see many features present between $P_{0}/P_{3} = 0.05$ and 0.35 cpp. In the side panel showing the horizontally collapsed spectrum, the peak at $P_{0}/P_{3} = 0.05$ cpp is the most prominent and we identify it as caused by the nulls. The next two peaks at 0.1 and 0.166 cpp we identify them as the main drift features ($P_{3} = 10\ P_{0}$ and $P_{3} = 6\ P_{0}$ respectively), whereas the pair of peaks visible at $P_{0}/P_{3} = 0.2$ and 0.33 cpp are their second harmonics. Inspection of the S2DFS plot (Fig.~\ref{fig:null-induced_mode_change}, left-hand plot) allows us to determine the ``A'' mode as being the most prominent. One can easily see the change from the ``A''-mode to ``B''-mode around block 1100. We also note that in first 700 blocks both drift modes are not easily seen in the S2DFS plot. We interpret this as being due to the frequent changes between drift modes, which resulted in a decrease of the spectral power of features corresponding to the drift modes. The S2DFS plot also explains the remaining features present around 0.1 cpp (e.g. block 400) as being created during the abrupt switch between the drift modes.

\section{Observations}

The above results clearly present the S2DFS technique as a compliment of the 2DFS method for detecting and analysing the drifting subpulses phenomenon, particularly for sources with multiple or changing drift rates. To test the new method using real observations, we have selected a number of archival data sets taken at the 94-m equivalent Westerbork Synthesis Radio Telescope (WSRT) in the Netherlands. The three selected pulsars, PSRs B0031$-$07, B1819$-$22 and B1944$+$17, were observed between 2000 and 2004 at two wavelengths of 21 and 92 cm. Specifically, we have used the PSR B0031$-$07 data from the work of \citet{smk05}, while the data of PSR B1819$-$22 and PSR B1944$+$17 were taken from \citet{wes06, wse07}. We have selected these sources based on the quality of data and the well known properties of these pulsars. This will allow us to compare the results from the 2DFS analysis with the new S2DFS technique. For a detailed discussion of the observing system and data reduction procedures see \citet{vkv02} and \citet{wes06}, while Table~\ref{table:pulsar_details} summarises the details of the observing sessions and results from the 2DFS analysis presented in \citet{wes06, wse07}.

\subsection{PSR B0031$-$07}

\begin{figure*}
\resizebox{\hsize}{!}{\includegraphics{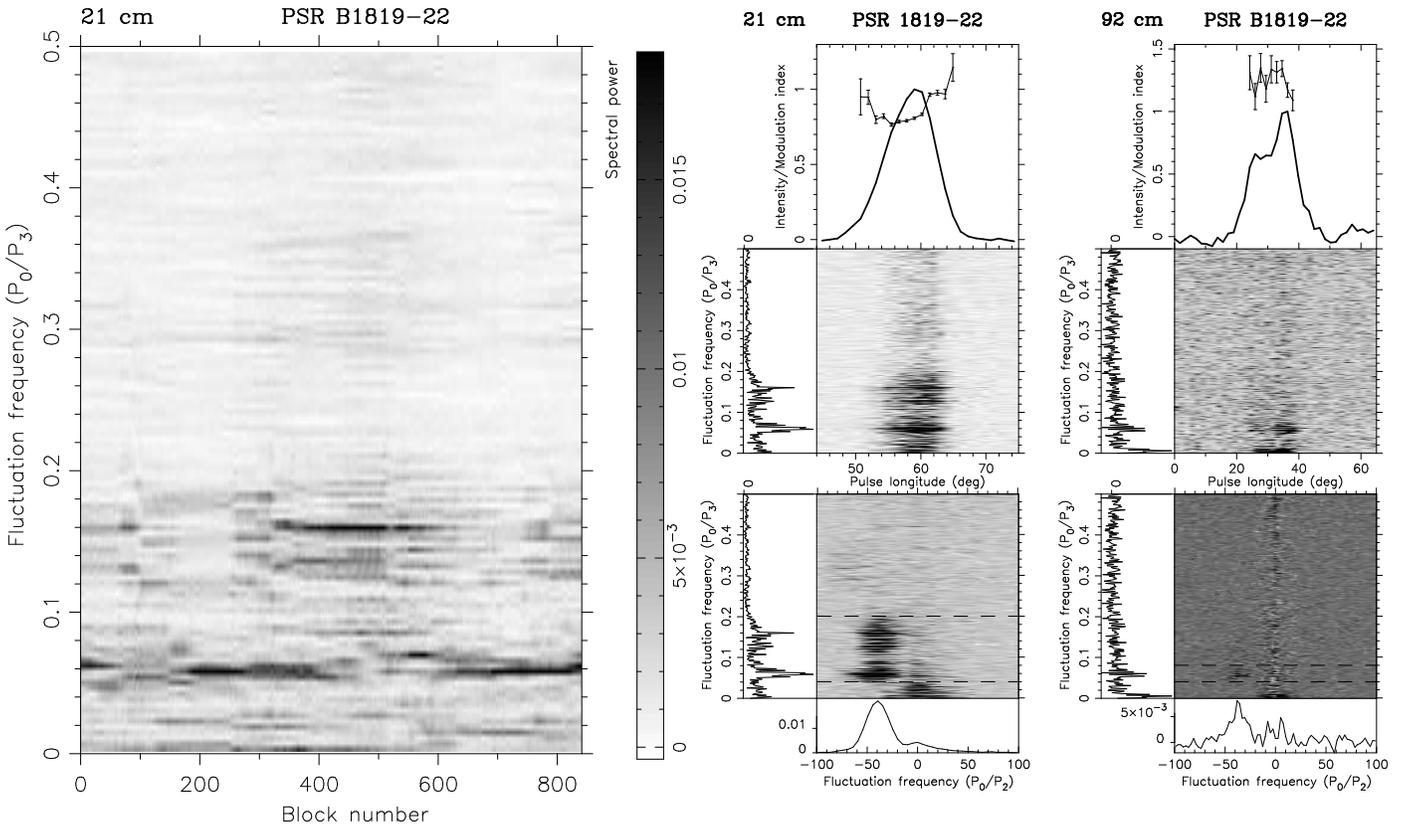}}
\caption{The analysis results shown for PSR B1819$-$22. For the explanation of the plots we refer to Fig.~\ref{fig:coherent_drifting} and the text.}
\label{fig:1819}
\end{figure*}

This pulsar shows a broad drifting feature in its 2DFS and S2DFS at 92 cm (Fig.~\ref{fig:0031}, left-hand and middle plot). In the case of the S2DFS plot we only show first 1500 blocks in order to keep the figure legible as the whole data set has 7411 blocks. Three drift modes have been found for this pulsar by \citet{htt70} at 145 and 400 MHz. In the work of \citet{wse07} most of the power in the 2DFS at 92 cm is due to the ``B''-mode drift ($P_{3} = 6\ P_{0}$) at 0.15 cpp. The slope of the drift bands changes slightly from band to band (e.g. \citealt{vj97}), causing the feature to extend vertically in the 2DFS plot. The results from the S2DFS analysis confirm the ``B''-mode drift as the most prominent one. One can see that the region, in the S2DFS plot, corresponding to the ``B''-mode drift seems to consist of three separate tracks. We interpret this as being caused by the changing slopes of the drift bands.

The ``A''-mode drift ($P_{3} = 12\ P_{0}$) is visible as a downward extension of the main drift feature to 0.08 cpp as can be seen in the 2DFS greyscale plot (Fig.~\ref{fig:0031}, bottom panel of middle plot). During the ``A''-mode the pulsar signal becomes weaker which can be directly seen in the pulse stack and as a decrease of spectral power of the feature in the S2DFS plot. In the case of the S2DFS an exact location and duration of the track corresponding to this drift mode can be seen in the left-hand plot of Fig.~\ref{fig:0031} between blocks 750 and 1000 at 0.08 cpp. Two more regions (not shown), denoting the presence of the ``A''-mode drift, were revealed after visual inspection of the whole S2DFS plot for this pulsar. We note the presence of a low-frequnecy feature, located close to $P_{0}/P_{3} = 0$ border. We identify both pulse-to-pulse modulations and nulling responsible for the presence of this feature as confirmed in the results from the analysis of simulated data. However, during the presence of the ``A''-mode drift the low-frequency feature disappears. We interpret this as follows: the region in the pulse stack corresponding to the presence of the slow drift mode shows no nulling as well as a decrease of the modulation index reaching the value of 1.3, which indicates low pulse-to-pulse modulation. The ``C''-mode drift at 0.25 cpp ($P_{2} = -19^{+2}_{-1}$ deg and $P_{3} = 4.1 \pm 0.1\ P_{0}$) reported by \citet{wse07} as seen in the 2DFS is also visible, although weakly, in the S2DFS plot (not shown). In both methods the feature at 0.3 cpp is the second harmonic of the main drift feature.

In the 21-cm data, \citet{wes06} report the ``A''-mode dominated over the ``B''-mode which is visible, although weakly, while the ``C''-mode was not detected in the 2DFS. In the case of results using the S2DFS (not shown for the 21 cm data) only the ``A''-mode is seen in the plot. The remaining ``B''- and ``C''-drift modes were not detected as the insufficient S/N makes it difficult to discriminate the features responsible for the drift from the noise in the S2DFS plot. The behaviour of this pulsar in the 92 cm and 21 cm observations is consistent with the multifrequency study of \citet{smk05, sms+07}.

\subsection{PSR B1819$-$22}

\begin{figure*}
\resizebox{\hsize}{!}{\includegraphics{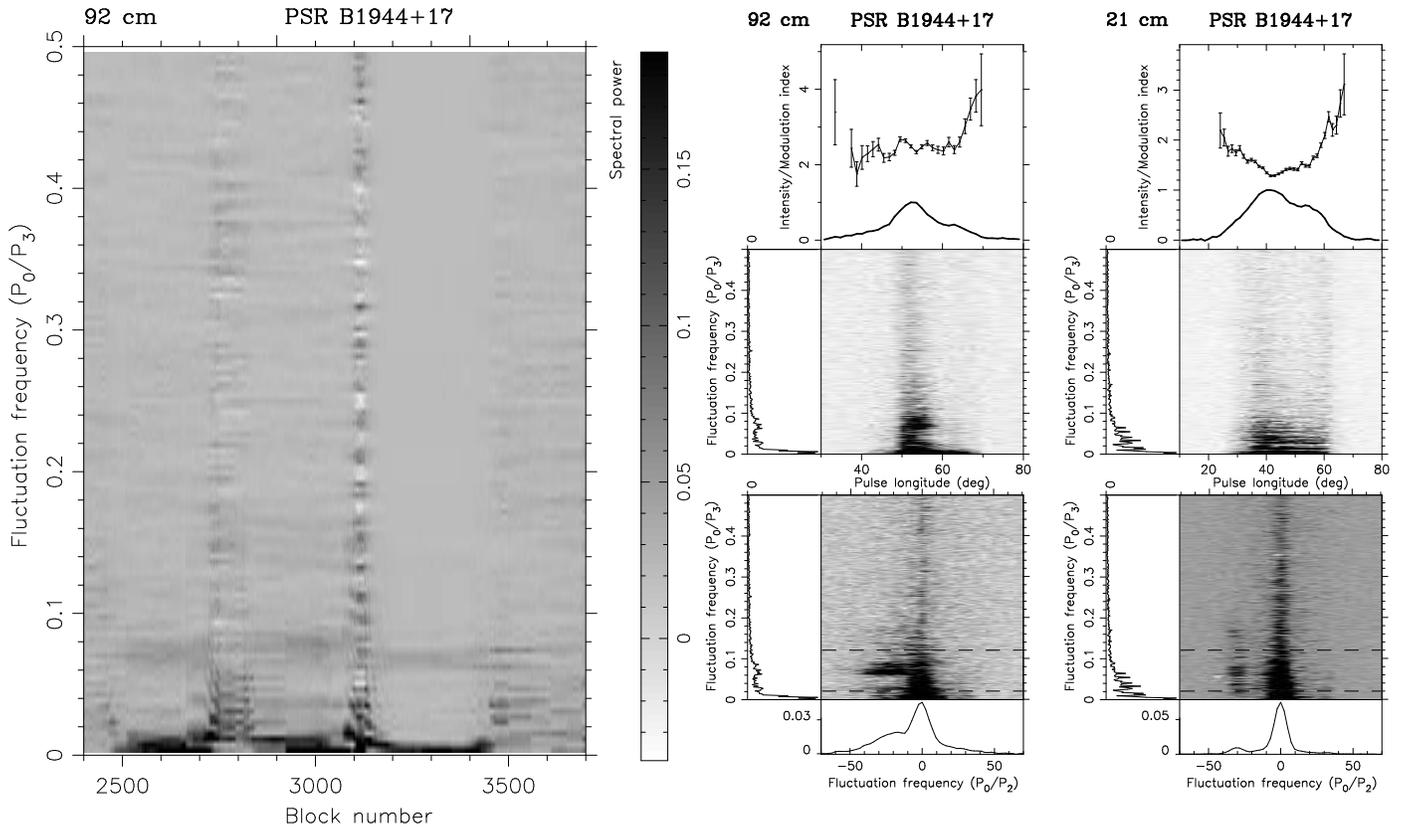}}
\caption{The analysis results shown for PSR B1944$+$17. For the explanation of the plots we refer to Fig.~\ref{fig:coherent_drifting} and the text.}
\label{fig:1944}
\end{figure*}

The results from the 21 cm data show a feature in the 2DFS plot (Fig.~\ref{fig:1819}, bottom panel of the central plot) which is double-peaked and corresponds to the two separate drift modes. The slower drift mode, present at 0.06 cpp and fast drift mode present at 0.16 cpp \citep{wes06}. In Fig.~\ref{fig:1819} (left-hand plot), the S2DFS plot clearly shows that the fast drift mode appears between blocks 300 and 600 at 0.16 cpp, corresponding to $P_{3} = 8 \pm 4\ P_{0}$ and $P_{2} = -50^{+110}_{-15}$ deg as reported by \citet{wse07} (see Table~\ref{table:pulsar_details}). We note that the visual inspection of the pulse stack revealed sparsely present fast mode drift bands in the whole pulse stack. In the case when only solitary drift bands are present one cannot see any contribution in the S2DFS, while just few consecutive drift bands can be resolved by the S2DFS as can be seen in Fig.~\ref{fig:1819} (left-hand plot, e.g. blocks 80 to 100). We also note that the drift mode change is preceded by a null, that can last from few down to a single pulse period and can be seen only after a visual inspection of the pulse stack.

We note several regions, in the S2DFS plot, where one cannot see clear tracks denoting drift (Fig.~\ref{fig:1819} (left-hand plot, e.g. blocks 100 to 170), we interpret this as either the presence of nulling or rapidly changing drift modes causing smearing out in the DFT because a too long transform length was used. However, introducing shorter transform length decreased the spectral resolution of the spectra and changed the S/N per spectral bin. This made the identification of features marking rapidly changing drift modes unattainable.

The 2DFS and S2DFS of this pulsar at 92 cm very clearly show a drift feature at 0.06 cpp (Fig.~\ref{fig:1819}, bottom panel of the right-hand plot) which corresponds to $P_{3} = 16.9 \pm 0.6\ P_{0}$ and $P_{2} = -17^{+2}_{-6}$ deg as reported by \citet{wse07}. This feature corresponds to the feature denoting the slow drift mode present in the 21 cm data while the fast drift mode is not present in the 92 cm data. There is a low-frequency excess present in the 2DFS plot (Fig.~\ref{fig:1819}, bottom panel of the right-hand plot) and it is stronger than in the 21 cm data. We interpret this as being due to the frequent nulls appearing in the pulse stack instead of a fast drift mode pulses.

\subsection{PSR B1944$+$17}

This pulsar shows drift features in its 2DFS at 92 and 21 cm (Fig.~\ref{fig:1944}, middle and right-hand plot) and the drifting can clearly be seen by eye in pulse stacks at both observed wavelengths. The features in the 2DFS plot are broadened because this pulsar shows drift-mode changes (\citealt{dchr86}, at both 430 and 1420 MHz). The $P_{3} = 13\ P_{0}$ ``A''-mode drift is visible in the 2DFS plot at 0.075 cpp at 92 and 21 cm.

The results from the S2DFS analysis for both wavelengths showed the presence of tracks in the S2DFS plot revealing the presence of the aforementioned drift mode. The sparse occurrence and very short duration times of the tracks (of the order of one hundred blocks) were caused by the frequent occurrence of nulls disrupting the drift bands. In the left-hand plot of Fig.~\ref{fig:1944} we show the S2DFS plot of 1300 out of 3735 blocks with the most prominent example of tracks denoting ``A''-mode drift present at 92 cm.

One can easily see that between block 3100 and 3150 there is a region where power in the S2DFS plot is dominated by fluctuations across the whole $P_{0}/P_{3}$ range. The visual inspection of the pulse stack revealed the presence of a group of pulses, which did not show any particular organised order, in between two long nulls disrupting the ``A''-mode drift. We interpret this as follows: The DFT window sliding through the long null, enclosed the group of pulses, and thus produced fluctuations across the $P_{0}/P_{3}$ range in the collapsed fluctuation spectra. One can also see that after block 3500 the low-frequency feature decreases its spectral power. We explain that as being caused by similar behaviour as in PSR B0031$-$07.

We also note that after this null the position of the track in the S2DFS plot denoting the ``A''-mode drift (Fig.~\ref{fig:1944}, left-hand plot, block 3150 to 3450) changes its position with respect to the track before the null. The 2DFS analysis of the pulse stacks containing the drift bands from the aforementioned region revealed that the drift rate value changes from $P_{3} = 13.5\pm 0.9\ P_{0}$ to $P_{3} = 14.3\pm 0.6\ P_{0}$. Without the use of the S2DFS it would not be possible to see that these changes are discrete changes in dirft rate and not smooth. It also shows that PSR B1944$+$17 belongs to the class of null-induced drift mode changing pulsars.

The $P_{3} = 6.4\ P_{0}$ ``B''-mode drift (0.16 cpp) did not appear in the 2DFS plot as a peak, although the centroid of the power was offset from the vertical axis by up to at least 0.2 cpp in the vertical direction at the wavelength of 21 cm (see Fig.~\ref{fig:1944}, right-hand plot) indicating its presence is either weak or only very sporadic. This drift mode was only visible in the S2DFS plot in the 21 cm observation. We note the low-frequency excess in both observations due to the presence of nulling.

\section{Summary and conclusions}

We have presented a new technique called the Sliding Two-Dimensional Fluctuation Spectrum (S2DFS), which is a good supplementary tool to the 2DFS and is capable of discovering and characterising the temporal changes of drifting subpulses from radio pulsars caused by e.g. null-induced mode or drift rate changes, something which cannot be done using the 2DFS method alone. We note that for the S2DFS method the S/N of the analysed data and length of the DFT window play a crucial role in obtaining sufficient quality of the maps of the collapsed fluctuation spectra. We have chosen the size of the DFT window to be 256 pulses which proved to be the best compromise between the resolution of the spectra (and hence the S2DFS plot) and sensitivity of the method to the short-lasting events like e.g. mode changes or nulls.

In order to test the S2DFS method we have generated a number of artificial pulse stacks. The emission models comprised of five scenarios: coherently drifting subpulses, coherently drifting subpulses with nulling, drift rate change, drift rate change with nulling and null-induced mode change. The set of pulse stacks generated for testing purposes comprised of the aforementioned different emission scenarios with additional effects like steady emission component or a pulse-to-pulse modulation introduced to the pulse stacks. These pulsar-intrinsic emission effects are known to disrupt drift bands and alter the results of spectral analysis. The maps of the collapsed fluctuation spectra allowed us to recover occurrence times of mode changes in the case of the null-induced mode change scenario and we were also able to resolve which one of the modes was more prominent. The simulated drift rate changes were easily revealed, something that could not have been done by the 2DFS only.

We have also used data sets from archival WSRT observations at wavelengths of 92 and 21 cm. We have analysed three pulsars, PSRs B0031$-$07, B1819$-$22 and B1944$+$17, which were selected based on the quality of the data and their known drift properties. All three sources are known to exhibit mode changes which is most prominent in the case of PSR B1819$-$22 at the wavelength of 21 cm and could be easily seen in the S2DFS plot. The usefulness of the new method is especially apparent in the case of short periods of different drift modes, where it is easier to find the moment of change, relative strength or coherence of the drift mode without having to visually inspect the pulse stack. Moreover, inspection of the pulse stack does not directly give the drift rate values. By applying this method to a large sample of pulsar such analysis would help to understand when and why drifting is non-coherent in different sources. We also note the potential of this method when applied to milisecond radio pulsars. If drift rate changes are discovered in recycled pulsars it would give significant insight into the pulse profile evolution of millisecond pulsars. This knowledge could be used for the construction of more precise timing models and hence, improving timing precision.

\begin{acknowledgements}
MS was supported by the EU Framework 6 Marie Curie Early Stage Training programme under contract number MEST-CT-2005-19669 ``ESTRELA''.
\end{acknowledgements}
\bibliographystyle{aa}

\end{document}